\documentclass{moriond}

\def\Journal#1#2#3#4{{#1} {\bf #2}, #3 (#4)}

\usepackage{amsmath}
\usepackage{amssymb}

\def\be{\begin{equation}}
\def\ee{\end{equation}}
\def\bea{\begin{eqnarray}}
\def\eea{\end{eqnarray}}

\def\gb{g_\text{bar}}
\def\go{g_\text{obs}}
\def\ag{\alpha_\text{grav}}

\newcommand{\Photo}{\includegraphics[height=35mm]{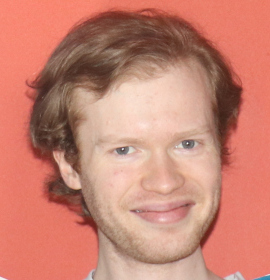}}

\begin{document}
\vspace*{4cm}
\title{Modified Newtonian Dynamics: Observational Successes and Failures}

\author{ Harry Desmond }

\address{Institute of Cosmology \& Gravitation, University of Portsmouth, Dennis Sciama Building,\\ Portsmouth, PO1 3FX, United Kingdom}

\maketitle\abstracts{
Modified Newtonian Dynamics (MOND) is an alternative to the dark matter hypothesis that attempts to explain the ``missing gravity'' problem in astrophysics and cosmology through a modification to objects' dynamics. Since its conception in 1983, MOND has had a chequered history. Some phenomena difficult to understand in standard cosmology MOND explains remarkably well, most notably galaxies' radial dynamics encapsulated in the Radial Acceleration Relation. But for others it falls flat---mass discrepancies in clusters are not fully accounted for, the Solar System imposes a constraint on the shape of the MOND modification seemingly incompatible with that from galaxies, and non-radial motions are poorly predicted. An experiment that promised to be decisive---the wide binary test---has produced mainly confusion. This article summarises the good, the bad and the ugly of MOND's observational existence. I argue that despite its imperfections it does possess ongoing relevance: there may yet be crucial insight to be gleaned from it.
}

\section{Introduction}\label{sec:intro}

There is abundant evidence that, when analysed using General Relativity (GR), there is much more mass in the universe than the measurable stars and gas. On the smallest scales, dwarf spheroidal galaxies have ``dynamical masses'' (inferred from their internal kinematics) $\sim$500$\times$ larger than their visible mass; on the largest, the primordial perturbations observed in the cosmic microwave background could not have formed the observed low-redshift large-scale structure without significantly faster growth than afforded by the universe's baryons. There are two possible solutions: either we retain GR and introduce more matter, or we retain the observed mass and enhance objects' dynamical response to it. The former---overwhelmingly the most popular---leads to the concept of dark matter and thence the standard model of cosmology, $\Lambda$CDM. The latter---the subject of this review---leads to MOND.

In 1983, Mordehai Milgrom hit upon a simple scheme for explaining galaxies' dynamics without dark matter.\cite{mi1,mi2,mi3} He proposed the following modification to a tracer's particle's dynamical acceleration $\vec{g}_\text{obs}$ produced by the Newtonian gravitational field sourced by the visible baryons, $\vec{g}_\text{bar}$:
\be\label{eq:mond}
\vec{g}_\text{obs} = \nu(g_\text{bar}/a_0) \: \vec{g}_\text{bar}
\ee
$\nu$ is called the interpolating function (IF), and any function will do for it provided it possesses the limits $\nu(x)\rightarrow1$ for $x\gg1$ and $\nu(x)\rightarrow x^{-1/2}$ for $x\ll1$. Classic choices are the ``Simple'' ($\nu(x) = 1/2 + (1/4 + 1/x)^{1/2}$), ``Standard'' ($\nu(x) = \frac{1}{\sqrt{2}} (1 + ((1 + 4/x^2)^{1/2})^{1/2})$), and ``RAR'' ($\nu(x) = 1/(1-\exp(-\sqrt{x}))$) functions.
$a_0$ is a new fundamental constant which marks the dividing line between the \emph{Newtonian regime} at high acceleration---where Newtonian dynamics is recovered---and the \emph{deep-MOND regime} at low acceleration where $\go \propto \gb^{1/2}$. The theory does not provide any way of calculating $a_0$ and hence it must be found empirically.

\emph{If} the modification to dynamics is based on acceleration,\footnote{\emph{That} the modification should be based on acceleration rather than some other dynamical variable is not clear \text{a priori}; its justification currently derives from the empirical correlations of galaxy properties.\cite{stiskalek}} the reason for the two limits to $\nu$ are clear: 1) Within the Solar System (SS; $\gb \gg a_0$) we must recover Newtonian mechanics where it has been precisely tested for centuries, and 2) in the outskirts of galaxies ($\gb \ll a_0$) we wish to produce flat rotation curves. Taking for simplicity circular motion around a spherically symmetric mass, this follows from
\be\label{eq:btfr}
\go=\frac{V^2}{r}, \:\:\: \gb=\frac{GM}{r^2} \:\: \Rightarrow \frac{V^2}{r} = \left(\frac{\gb}{a_0}\right)^{-1/2} \gb = \sqrt{a_0\: \gb} \: \: \Rightarrow V^4=a_0 G M
\ee
where $G$ is Newton's constant and $M$ the galaxy's total baryonic mass. When all of this mass is enclosed, the rotation velocity therefore becomes independent of galactocentric radius in the deep-MOND regime. Matching observed galaxy rotation velocities requires $a_0 \approx 1.2\times10^{-10}$ m/s$^2$, which, strangely enough, is very similar to some fundamental quantities in cosmology: $a_0 \approx c H_0 \approx c^2 \Lambda^{1/2}$, where $H_0$ is the present-day expansion rate and $\Lambda$ the dark energy scale.

Eq.~\ref{eq:mond} is little more than a fitting function. How to embed it in a full theory? The first thing to note is that it may be interpreted either as a modification to gravity---altering the gravitational field produced by a given mass---or as a modification to inertia, altering the response of an object to a given force. These lead to subtly different predictions in general, but the modified inertia idea has received very little attention and has so far resisted incorporation into a viable, fully-fledged theory.\cite{milgrom_mi} In contrast the modified gravity idea is straightforward to instantiate in theories that derive from Lagrangians, satisfy conservation laws and admit relativistic extensions. There are two classic nonrelativistic theories of MOND. The first, called AQUAL (for ``AQUAdratic Lagrangian''), alters the Poisson equation for the gravitational potential produced by matter:\cite{aqual}
\be
\nabla \cdot \left[\mu(|\nabla \phi|/a_0) \nabla \phi\right] = 4\pi G \rho_\text{bar}
\ee
where $\mu$ is defined by $\nu(x \mu(x)) = 1/\mu(x)$.
The second, called QUMOND (for ``QUasi-linear MOND''), retains the regular Poisson equation for producing the Newtonian potential $\phi_\text{N}$ from the density, but then introduces a new relation to map this to the gravitational potential that determines objects' dynamics:\cite{qumond}
\be
\nabla^2 \phi = \nabla \cdot \left[\nu(|\nabla \phi_\text{N}|/a_0) \nabla\phi_\text{N}\right]
\ee
The theoretical underpinnings of MOND are not the subject of this article; the reader is directed to comprehensive reviews\cite{sanders_mcgaugh,famaey_mcgaugh,banik_zhao} and the currently most sophisticated relativistic formulation.\cite{aest}

\section{The Good}\label{sec:good}

MOND produces flat rotation curves by construction. What else is it good for? Eq.~\ref{eq:btfr} describes a relation between the baryonic mass of a galaxy and its asymptotic rotation velocity: a power law with slope 4, zero intrinsic scatter and normalisation set by $a_0$. Although unknown observationally at MOND's inception, this relation was first measured in 2000 as the \emph{Baryonic Tully--Fisher Relation} (BTFR).\cite{btfr} Estimates of the slope and intrinsic scatter vary somewhat, but are typically 3.5-4 and $\lesssim 0.1$ dex respectively, in good agreement with the MOND prediction. In $\Lambda$CDM the BTFR must arise from the complex process of hierarchical galaxy formation: the \textit{a priori} assumption $M_\text{bar} \propto M_\text{halo}$ produces a slope of 3 (clearly discrepant with the data), and more recent abundance matching-based models produce an unobserved curvature and considerable intrinsic scatter.\cite{desmond_btfr} The fact that the BTFR residuals do not correlate with galaxy size, while following naturally from Eq.~\ref{eq:btfr}, is also nontrivial in $\Lambda$CDM.\cite{desmond_residuals}

Eq.~\ref{eq:mond} specifies not only the asymptotic rotation velocity but the entire rotation curve as a function of a galaxy's baryonic mass distribution, even when the Newtonian contribution to the rotation curve is much less than that observed. This provides a strong test of the theory as there are many hundreds of local galaxies with both well-measured baryon distributions and kinematics, and again the theory succeeds.\cite{sanders_mcgaugh} In $\Lambda$CDM it is highly unexpected that galaxies' dynamics (hence their dark matter) should be (near-)perfectly calculable from their baryonic mass on a galaxy-by-galaxy basis with just a single global parameter.

\begin{figure}
\begin{minipage}{0.5\linewidth}
\centerline{\includegraphics[width=1\linewidth]{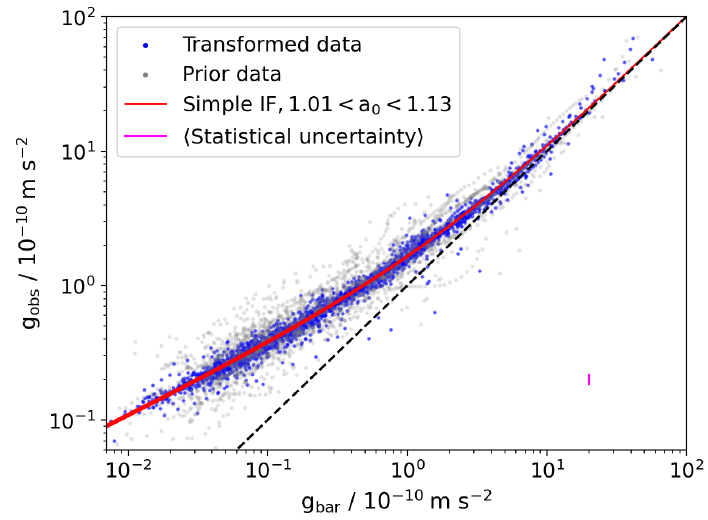}}
\end{minipage}
\hfill
\begin{minipage}{0.5\linewidth}
\centerline{\includegraphics[width=1\linewidth]{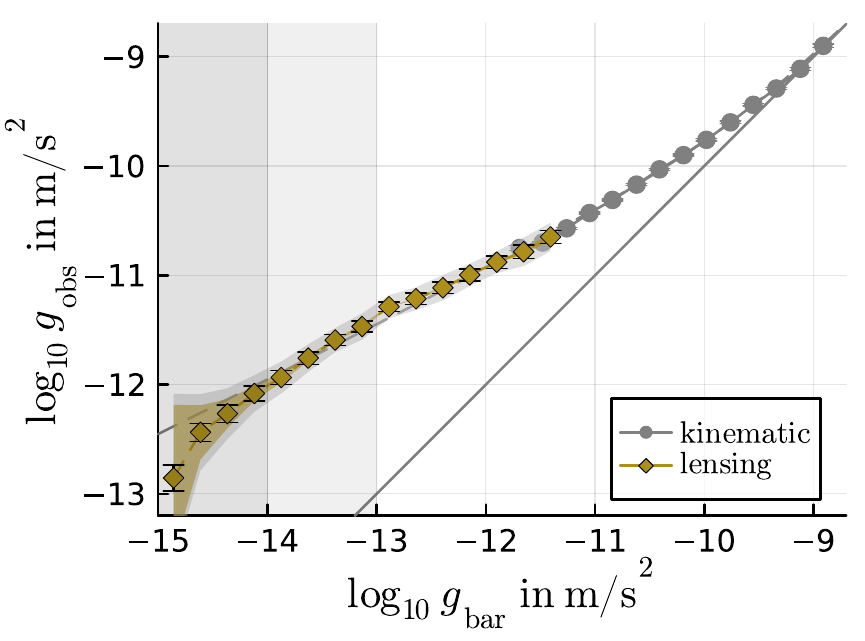}}
\end{minipage}
\caption[]{The radial acceleration relation. \emph{Left:} The RAR of the SPARC survey at maximum \textit{a priori} (grey) and \textit{a posteriori} (blue) parameter values after optimising the galaxy-wide nuisance parameters.\cite{urar} The average statistical uncertainty in the latter case is shown by the pink bar in the lower right; the 1-1 line is in dashed black and the best-fit Simple IF in red. \emph{Right:} The RAR obtained from stacked weak lensing in the KIDS survey.\cite{mistele} The grey points show the binned SPARC RAR and the shaded grey bands at $\gb<10^{-13}$ m/s$^2$ where the isolation criterion may fail (and hence the recovered $\go$ not perfectly describe the dynamical acceleration of the lenses).}
\label{fig:RAR}
\end{figure}

This success suggests the possibility of generalising the BTFR to link baryonic mass with kinematics across rotation curves. Such a relation was first studied in 1990 under the name of the mass discrepancy--acceleration relation,\cite{sanders_1990} and was revamped in 2016 as the radial acceleration relation (RAR).\cite{rar} This directly correlates the acceleration sourced by the baryons $g_\text{bar}$ with the total dynamical acceleration $g_\text{obs}$. With many such measurements per galaxy, the RAR is more powerful statistically than the BTFR which has only one. The RAR is currently best measured using the SPARC sample;\cite{sparc} the grey points in Fig.~\ref{fig:RAR} (left) show the fiducial SPARC RAR for 2696 measurements from 147 late-type galaxies.\cite{rar}

The RAR is a direct manifestation of Eq.~\ref{eq:mond}. This predicts a power-law slope of 1 at high $g_\text{bar}$ and $1/2$ at low $g_\text{bar}$, and negligible intrinsic scatter, all of which are (at least approximately) realised observationally.\cite{rar,sr-rar} None of these properties are expected (at least \textit{a priori}) in $\Lambda$CDM, where again the relation must emerge as a consequence of a chaotic galaxy formation procedure. That $g_\text{obs} \rightarrow g_\text{bar}$ as $g_\text{bar} \rightarrow \inf$ is at least plausible (indicating that the central regions of high-surface-brightness galaxies are baryon-dominated), but there is no known reason why the RAR should have the shape it does as low $g_\text{bar}$, why it should be so tight, why there should exist a ``characteristic acceleration'' $a_0$---or indeed why $g_\text{bar}$ and $g_\text{obs}$ should be the right variables to correlate in the first place.\cite{desmond_mdar}

What exactly is the relation's intrinsic scatter? Simply fitting the SPARC data with the Simple IF, assuming their fiducial error model and Gaussian, uncorrelated uncertainties, yields $0.082 \pm 0.003$ dex.\cite{urar} However, most of the scatter in the points and the sizes of the errorbars derive from uncertainties in galaxy properties that link $g_\text{bar}$ and $g_\text{obs}$ across particular galaxies' rotation curves, viz. their mass-to-light ratios, distances and inclinations. The proper procedure is to constrain these galaxy nuisance parameters simultaneously with the parameters of the RAR and then recalculate $g_\text{bar}$ and $g_\text{obs}$ according to the best-fitting parameters. This produces the blue points in Fig.~\ref{fig:RAR} (left),\cite{urar} which possess the extraordinarily small intrinsic scatter of $0.034\pm0.002$ dex and no correlated deviations from the Simple or RAR IF fit. And this is not some artifact of the SPARC sample: a new, completely independent calibration from the MIGHTEE survey makes the intrinsic scatter $0.045\pm0.022$ dex.\cite{andreea}

The extreme tightness and regularity of the RAR suggests the question of whether it is somehow special among the set of dynamical correlations of late-type galaxies. This is an immediate consequence of MOND, but would be bizarre---perhaps even inexplicable---in $\Lambda$CDM. Four properties were recently proposed that any dynamical relation would have to have to be considered fundamental:\cite{stiskalek} 1) it cannot have significant residual correlations with any other variable, 2) it must be the tightest projection of galaxies' dynamical parameter space, 3) it must be capable of accounting for all other dynamical correlations in conjunction with the non-dynamical correlations present in the dataset in question, and 4) it must be unique in having these properties. Extraordinarily, the RAR satisfies all four.\cite{stiskalek}

The RAR is a phenomenon not only of late-type galaxies: early-types also follow the relation, albeit with larger observed scatter.\cite{rar} A more fundamental extension of the RAR however comes from replacing the kinematic tracers of $g_\text{obs}$ (the velocities of stars and gas) with a measure of dynamical acceleration derived from stacked weak lensing.\cite{brouwer,mistele} This can extend the RAR to much lower accelerations. Fig.~\ref{fig:RAR} (right) shows the relation so measured in the KIDS survey: the RAR appears to continue unchanged to exceedingly low $g_\text{bar}$. This is equivalent to rotation curves remaining flat out to $\sim$1 Mpc, which would not be expected in $\Lambda$CDM: these distances probe near the halos' virial radii where the standard Navarro--Frenk--White density profile falls off as $1/r^3$, leading to a declining predicted rotation curve. The fall-off measured at very low $\gb$ may be a manifestation of the \emph{External Field Effect} (EFE) in MOND, whereby dynamics resumes Newtonian scalings when the external field of an object, sourced by surrounding mass, exceeds the internal field.
Although something resembling the RAR arises in $\Lambda$CDM,\cite{desmond_mdar} it seems quite unlikely to possess all the remarkable properties of the observed relation.

While the RAR is (in the author's opinion) the best evidence for MOND, it is not the only. Others, which there is not space here to describe, include a possible tidal dwarf solution to the planes of satellites problem,\cite{kroupa_dual_dwarf} a lack of dynamical friction in galaxy bars,\cite{roshan} and a putative direct detection of the EFE.\cite{chae_efe}

\section{The Bad}\label{sec:bad}

Is MOND a magic bullet for our astrophysical problems? By no means.

\begin{figure}
\begin{minipage}{0.5\linewidth}
\centerline{\includegraphics[width=1\linewidth]{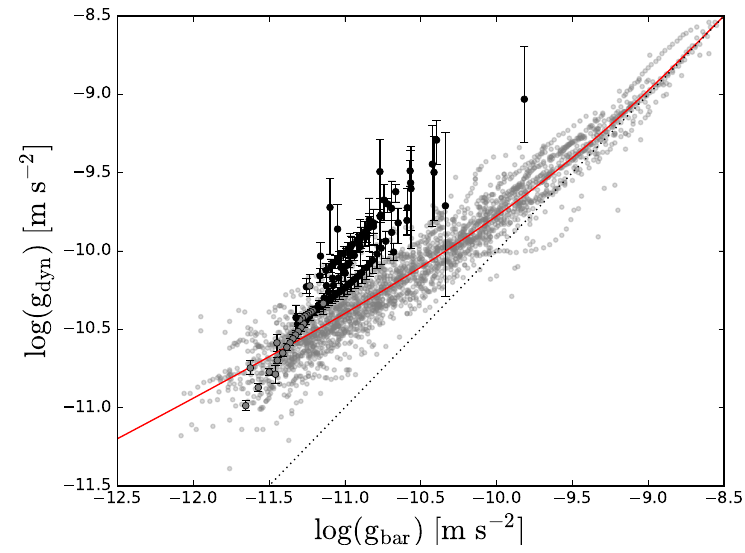}}
\end{minipage}
\hfill
\begin{minipage}{0.5\linewidth}
\centerline{\includegraphics[width=0.8\linewidth]{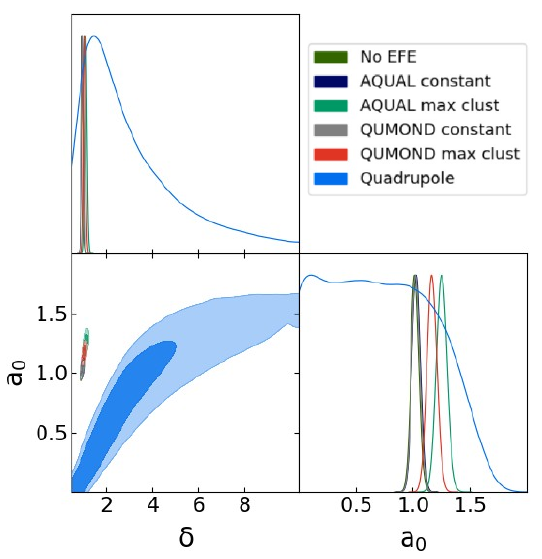}}
\end{minipage}
\caption[]{\emph{Left:} The RAR of 10 galaxy clusters (black and dark grey points).\cite{Li} The light grey points show the prior SPARC RAR (as in Fig.~\ref{fig:RAR}, left). \emph{Right:} Constraints on $a_0$ and the shape parameter $\delta$ from the Cassini $Q_2$ constraint in blue, and from the SPARC RAR under various models for the EFE.\cite{cassini} The discrepancy between the RAR and quadrupole constraints is $9\sigma$.}
\label{fig:bad}
\end{figure}

MOND does not appear to work at either larger or smaller scales than that of galaxies. The longest-known problem is that using the same value of $a_0$ as for galaxies does not fully explain clusters' dynamical masses. The missing mass goes down, but is not eliminated.\cite{sanders_mcgaugh} This implies either a stronger MOND boost in clusters,\cite{emond} or actual missing mass. A consequence of this phenomenon is that in merging clusters (most notably the Bullet Cluster\cite{bullet_cluster}), much of the lensing mass is unaccounted for in MOND, and, awkwardly for the theory, appears to trace the collisionless galaxies (where the dark matter would be expected to be) rather than the bulk of the baryonic mass in the X-ray emitting gas. In fact, however, the MOND boost is not always not enough in clusters; sometimes it is too much, as evidenced by points lying \emph{below} the galaxy RAR (see Fig.~\ref{fig:bad}, left).\cite{Li} For this to be explained with missing mass, that mass would have to be negative. A distinct but still cluster-related problem is that ultra-diffuse galaxies within the Coma cluster have been observed to have velocity dispersions consistent with the isolated MOND prediction, but not, as would be expected, when including the impact of the EFE.\cite{Freundlich}

Small scales are also problematic. We've seen that the radial dynamics of disk galaxies---epitomised by the RAR---works excellently in MOND, but what about motions perpendicular to the disk? This has been explored by comparing MOND and cold dark matter models for the Milky Way.\cite{Lisanti} With a suitable choice of halo properties, either can account for the rotation curve. For a given radial force, MOND predicts a stronger vertical force because the gravity is sourced entirely by the thin baryonic disk rather than a spherically symmetric halo. This can be tested by measuring the vertical velocity dispersion, which, for realistic properties of the MW disk, agrees with the dark matter but not MOND prediction.

An even more significant problem lies closer to home. At the location of the SS, the gravitational field of the MW has a strength of 1.8$\:a_0$. This suppresses the MOND boost much below that of the deep-MOND regime, but residual effects remain. In particular, the external field, pointing towards the Galactic centre, introduces a quadrupole into the SS's gravitational field which perturbs planetary orbits. Saturn's trajectory should be altered sufficiently to have been measurable by the Cassini spacecraft which orbited the planet from 2004 to 2017. The amplitude of the quadrupole is denoted $Q_2$, and was constrained by Cassini to be $Q_2 = 3 \pm 3 \times 10^{-27}$ s$^{-2}$.\cite{hees_1} This agrees with the Newtonian expectation of 0, but classic MOND models predict $\sim$35.\cite{hees_2}

MOND's $Q_2$ prediction depends crucially on the shape of the IF as it transitions from the deep-MOND to Newtonian regimes. The steeper the rise, the closer it is to Newtonian behaviour at 1.8$\:a_0$ and hence the lower the predicted $Q_2$. The theory \textit{per se} has no problem with this: the transition could even be a step at $a_0$. However, the shape of the IF is constrained also by the RAR, which maps out the transition region. This may be quantified by fitting both the SS quadrupole and SPARC RAR by ``families'' of IFs containing a shape parameter specifying the sharpness of the MOND--Newton transition. An example is the ``delta-family'', $\nu(x) = (1-\exp(-x^{\delta/2}))^{-1/\delta}$, which recovers the RAR IF at $\delta=1$. The constraints on $a_0$ and $\delta$ are shown in Fig.~\ref{fig:bad} (right) for various models for the EFE in the RAR:\cite{cassini} the RAR requires a shallower MOND--Newton transition than required by the Cassini measurement at $\sim$9$\sigma$. This can be ameliorated somewhat by relaxing the SPARC prior on $K$-band mass-to-light ratio, and even more so by removing galaxies with bulges, but these are \textit{ad hoc} changes without theoretical justification. The conclusion that the SS is more Newtonian than MOND expects is reinforced by the result that Newtonian but not MONDian simulations can account for the distribution of binding energies of long-period and Oort-cloud comets.\cite{comets} Further RAR studies are needed to determine whether galaxy dynamics really is incompatible with a sharp ($\delta\gtrsim2-3$) IF.

\section{The Ugly}\label{sec:ugly}

A crucial test of MOND is the wide binary test (WBT). The WBT examines the kinematics of stars in binaries in the Solar neighbourhood, specifically those with wide enough separation for their internal acceleration to be in the MOND regime.
Absent the EFE from the MW, this would produce rotation velocity independent of the stars' separation. Including the EFE, classic MOND models would predict Keplerian trajectories with a $\sim$20\% boost to orbital velocities.\cite{pittordis_sutherland}

\begin{figure}
\begin{minipage}{0.5\linewidth}
\centerline{\includegraphics[width=1\linewidth]{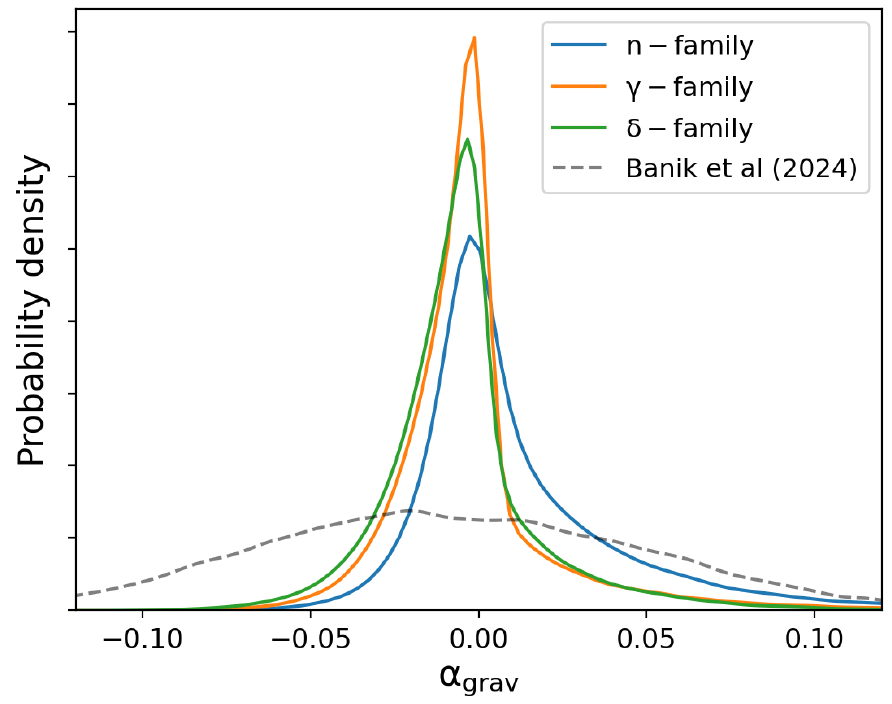}}
\end{minipage}
\hfill
\begin{minipage}{0.5\linewidth}
\centerline{\includegraphics[width=1\linewidth]{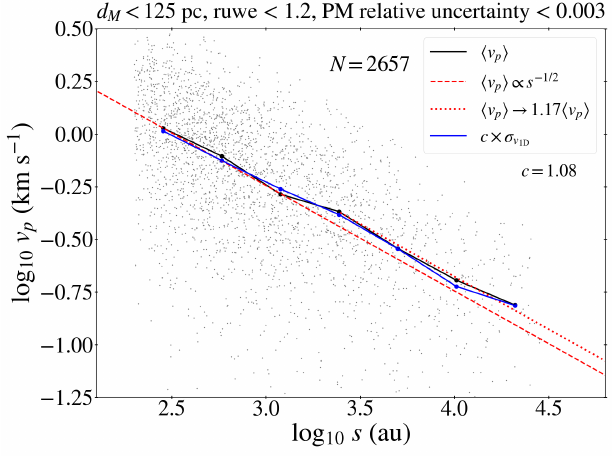}}
\end{minipage}
\caption[]{\emph{Left:} The posterior predictive distribution from the $Q_2$ constraint (blue contour of Fig.~\ref{fig:bad}, left) for the parameter $\ag$ that summarises the outcome of the WBT, for three IF families, with the posterior from one WBT analysis using \textit{Gaia} DR2 overlaid in dashed grey.\cite{cassini} These results confidently indicate a Newtonian SS and Solar neighbourhood. \emph{Right:} Projected velocity versus binary separation from an alternative WBT using the same data,\cite{Chae_1} apparently indicating a MONDian transition at high separation. This would imply $\ag\approx1$.}
\label{fig:WBT}
\end{figure}

Just like the SS quadrupole, the WBT is sensitive to the steepness of the MOND IF. If it is sufficiently steep that Newtonian behaviour is recovered at 1.8$\:a_0$, MOND would produce no kinematic boost. Since we have already constrained this steepness from the Cassini $Q_2$ measurement, we can make a prediction for the outcome of the WBT. This outcome may be summarised by a parameter $\alpha_\text{grav} \equiv \frac{\sqrt{\eta}-1}{0.193}$,
where $\eta \equiv \langle g_r \rangle / g_\text{N}$ is the excess of the azimuthally averaged asymptotic radial gravity over the Newtonian prediction.
This is defined such that $\ag=0$ describes Newtonian gravity and $\ag=1$ a fiducial MOND model (QUMOND with the Simple IF and $a_0=1.2\times10^{-10}$ m/s$^2$). The solid coloured lines in Fig.~\ref{fig:WBT} (left) show the posterior predictive distributions for $\ag$ from various IF-family fits to the $Q_2$ constraint: since this requires the SS to be all-but Newtonian, it forces WBs, at a similar acceleration, to be Newtonian too. Conversely, the RAR fits of Fig.~\ref{fig:bad} (right) predict $0.8 \lesssim \ag \lesssim 0.9$.

Which of these predictions agrees with the actual results of the WBT? Here's where the ugliness appears: the test has been conducted effectively four times, and the results are roughly uniformly distributed between Newton and MOND.

A recent analysis, using \textit{Gaia} DR3, obtains the constraint $\ag=-0.021^{+0.065}_{-0.045}$,\cite{banik_wbt} shown as the dashed grey posterior in Fig.~\ref{fig:WBT} (left). This is consistent with Newtonian mechanics and the Cassini constraint, and discrepant with the fiducial MOND model at 16$\sigma$. Prior to that,\cite{pittordis_sutherland} an analysis using \textit{Gaia} EDR3 claimed instead \emph{weak} evidence for Newtonian dynamics. The authors declined to draw a stronger conclusion because not even their best-fitting models provide a good fit to the data in an absolute sense, suggesting unknown systematics that may also invalidate the relative preference for Newton. This is in fact also true for the first analysis mentioned, urging caution in interpreting its apparently decisive rejection of MOND. The quality of the fits has been somewhat improved recently,\cite{pittordis_sutherland_new} offering hope that a successful model is on the cards.

But then we have studies instead \emph{preferring} MOND. Such results go back to the inception of the WBT. The first ``anomaly'' identified in the kinematics of WBs found the radial velocity to become independent of separation large separation (i.e. low acceleration).\cite{Hernandez_1} This is as expected from MOND \emph{without} the EFE, a highly unexpected scenario even within the MOND paradigm. The key result of a \emph{fourth} analysis is shown in Fig.~\ref{fig:WBT} (right),\cite{Chae_1} which indicates a Keplerian decline in rotation velocity but with a boosted amplitude as the acceleration crosses $a_0$, as one would expect from MOND \emph{including} the EFE. The claimed falsification of Newtonian behaviour is $5.8\sigma$.\cite{Chae_2} This corresponds to $\ag\approx1$, in clear disagreement with the $Q_2$ constraint. Were this the true WBT result, an explanation would therefore be required as to why dynamics is Newtonian within the SS but MONDian in the Solar neighbourhood (where the WBs live), despite these two environments occupying almost identical accelerations. This cannot be achieved by simply modifying the IF shape.\cite{cassini}

Why do these analyses find opposing results? There are several WBT systematics that are currently very difficult to account for robustly, including unseen extra stars within the binaries, ellipticity distributions and chance line-of-sight alignments. If one's treatment of these is insufficiently flexible, model misspecification may cause significant bias despite the constraints possessing high statistical precision. Other points of contention include treatment of uncertainties in the data, the quality cuts used to select the sample and how far into the Newtonian regime the sample extends. It is an uncomfortable fact for proponents of MONDian WBs that almost any form of unmodelled contamination will enhance the dynamics and hence masquerade as a MONDian signal. The author hopes that this controversy will increase scrutiny of the WBT in the future, ideally with a blinded methodology tested extensively on mock data.

\section{Summary and Conclusion}\label{sec:conc}

MOND is an interesting idea for doing away with dark matter by subtly altering dynamics at low accelerations. Its many successes during its 42-year life could not have been anticipated by $\Lambda$CDM. Its principal triumph is the prediction of a tight, regular and universal radial acceleration relation that plays a fundamental role in late-type galaxy dynamics. However, its powers do not extend far beyond galaxies. On smaller scales, the trajectory of Saturn and energies of long-period comets imply that the SS is far more Newtonian than MOND would predict, and vertical motions in disks conform better with the dark matter expectation. On larger, cluster, scales there is still missing mass on average (although not always), ultra-faint satellites evince a mysterious absence of the EFE, and the Bullet Cluster shows the dynamical mass to trace the collisionless galaxies rather than the far more massive collisional hot gas. Cosmology is another can of worms entirely: while MOND \textit{per se} does not come with a cosmology---and hence cannot obviously make contact with the many and varied successes of $\Lambda$CDM in that regard---there do exist relativistic parent theories that can recover the cosmic microwave background power spectrum and standard growth of structure.\cite{aest} The wide binary test, touted as a ``crossroads experiment'', has produced almost as many different results as people studying it.

What are we to make of this? I offer two suggestions for attempting to retain MOND's successes while circumventing its failures. First, it will be noted that (with the exception of clusters) MOND's failures occur for higher-frequency motions than those for which it succeeds. This suggests the possibility of introducing a frequency scale $\omega_0$ such that Newtonian dynamics is recovered if \emph{either} $a>a_0$ \emph{or} $\omega > \omega_0$. This may be natural behaviour in more sophisticated modified gravity MOND theories such as Generalised QUMOND.\cite{gqumond} Similar phenomenology could perhaps be obtained by equipping MOND with a screening mechanism, as is common practice to hide the fifth forces produced by scalar--tensor theories of gravity (that extend rather than fundamentally alter GR) in the SS. Second, as mentioned in Sec.~\ref{sec:intro}, modified inertia theories of MOND have been far less developed than modified gravity. There are in fact at least two empirical reasons for preferring modified inertia: 1) it produces the algebraic MOND relation (Eq.~\ref{eq:mond}) exactly in the RAR, as seems to be evidenced by Fig.~\ref{fig:RAR} (modified gravity theories predict some deviation from it), and 2) it can produce different effective IFs in different systems, perhaps reconciling the otherwise discrepant results that we have seen.\cite{milgrom_mi} In particular, one wonders if MOND might have something to do with Mach's principle. Could a change to dynamics at accelerations below ``that of the universe'' ($a_0 \approx cH_0 \approx c^2\Lambda^{1/2}$) reflect a transition in the way in which distant masses determine an object's inertia\cite{sciama,palomo}?

\end{document}